\documentclass[a4paper,11pt]{article}
\usepackage{pos}
\usepackage{cleveref}
\usepackage[T1]{fontenc}
\usepackage{xspace}
\usepackage{wrapfig}
\usepackage{subcaption}
\usepackage{graphbox}

\newcommand{\xmax}{\ensuremath{X_{\mathrm{max}}}\xspace}%

\title{Anisotropies, large and small}
 \ShortTitle{Anisotropies, large and small}

\author[a, b]{Teresa Bister}
\author*[c]{Glennys Farrar}

\affiliation[a]{Institute for Mathematics, Astrophysics and Particle Physics, Radboud University Nijmegen\\Nijmegen, The Netherlands}

\affiliation[b]{Nationaal Instituut voor Kernfysica en Hoge Energie Fysica (NIKHEF), \\Science Park, Amsterdam, The Netherlands}

\affiliation[c]{Center for Cosmology and Particle Physics, New York University\\New York, NY10003, USA}

\emailAdd{teresa.bister@ru.nl}
\emailAdd{gf25@nyu.edu}

\abstract{We report on several new results using anisotropies of UHECRs. We improve and extend the work of Ding, Globus and Farrar, who modeled the UHECR dipole assuming sources follow the dark matter distribution, accounting for deflections in the Galactic and extragalactic magnetic fields but using a simplified treatment of interactions during propagation. The work presented here employs an accurate and self-consistent treatment of the evolution of composition during propagation, allows for and explores the impact of “bias” in the relation between UHECR sources and the dark matter distribution, and investigates the possible generation of arrival-direction-dependent composition anisotropies. Limits on the source number density consistent with the observed anisotropies are derived for the case where UHECR sources follow the dark matter distribution, and compared to a homogeneous source distribution case.}

\ConferenceLogo{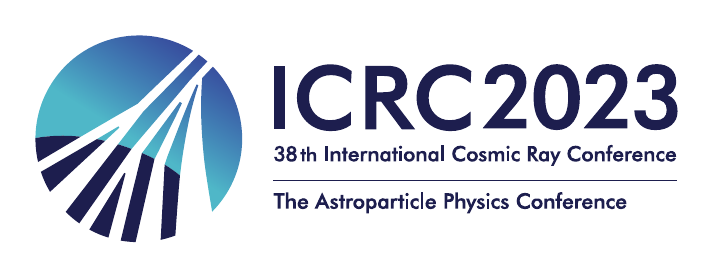}

\FullConference{%
38th International Cosmic Ray Conference (ICRC2023)\\
  26 July - 3 August, 2023\\
  Nagoya, Japan}

\begin{document}
\maketitle

\section{Introduction}
Recently, the Pierre Auger collaboration (hereafter \textit{Auger}) has published the significant discovery of a large-scale anisotropy of the ultra-high-energy cosmic ray (UHECR) arrival directions above 8 EeV~\cite{Auger_dipole_2017, Auger_dipole_2018}. The observed dipole anisotropy exhibits an amplitude of approximately 6\%, increasing with the energy. Its direction is not pointed toward the Galactic center, suggesting an extragalactic origin of UHECRs at these energies. The amplitude and directional characteristics of UHECR anisotropies can provide valuable insights into the possible distribution of UHECR sources, where the complex interplay between interactions with background photon fields and deflections by cosmic magnetic fields along the propagation from sources to Earth has to be considered. A scenario where the sources of UHECRs follow the Large-Scale Structure (LSS) of (dark) matter in the nearby universe was investigated before by C. Ding, N. Globus, and G. Farrar~\cite{Ding_2021} with promising results. We build upon the foundation of that work and introduce improvements regarding a proper treatment of propagation effects, a common injection at the sources, and a possible inclusion of a bias between UHECR sources and the LSS. Additionally, we investigate further anisotropy measures like the power spectrum and possible composition anisotropies, and set limits on the source density using these measures. Finally, we compare the model where the UHECR sources follow the LSS to a model where they are just randomly distributed. Note that similar ideas have also been investigated in~\cite{Allard_dipole_2021}, but without a fit of the injection parameters and with a different treatment of propagation effects. Additionally, the nearby source distribution in~\cite{Allard_dipole_2021} is explicitly sub-sampled from the flux-limited 2MRS catalog, while we use \textit{CosmicFlows 2}~\cite{Hoffman_CF} where the 3d source density is calculated from peculiar galaxy velocities. To avoid radial selection biases due to the flux-limited approach, in~\cite{Allard_dipole_2021} they have to set a lower limit on the source luminosity, limiting the range of possible number densities that can be explored to $n \lesssim 7.6 \times 10^{-3}$ / Mpc$^3$. We will show that larger number densities than that are in good agreement with Auger data, and that the UHECR production density can be approximated well by the CosmicFlows matter density without a bias.

\section{UHECR source model and fit method}
\begin{figure}[ht]
    \centering
    \begin{minipage}[b]{0.44\linewidth}
        \centering
        \includegraphics[width=\textwidth]{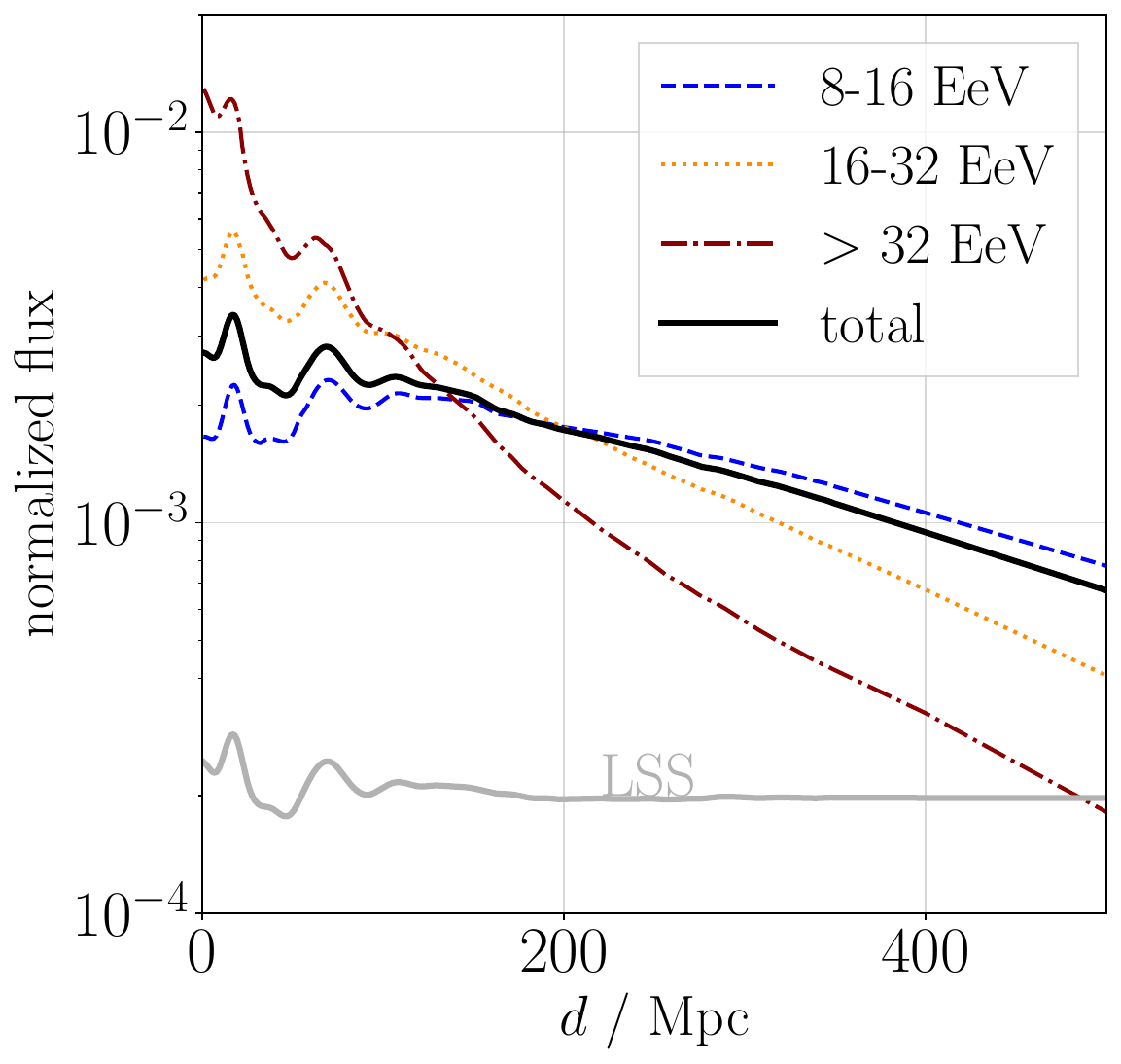}
        \caption{Flux contribution of different distances for three energy bins in the best-fit model. The distribution of the LSS (a.u.) is also shown in grey.} \label{fig:propagation}
    \end{minipage}
    \hfill
    \begin{minipage}[b]{0.54\linewidth}
        \centering
        \includegraphics[width=0.49\textwidth]{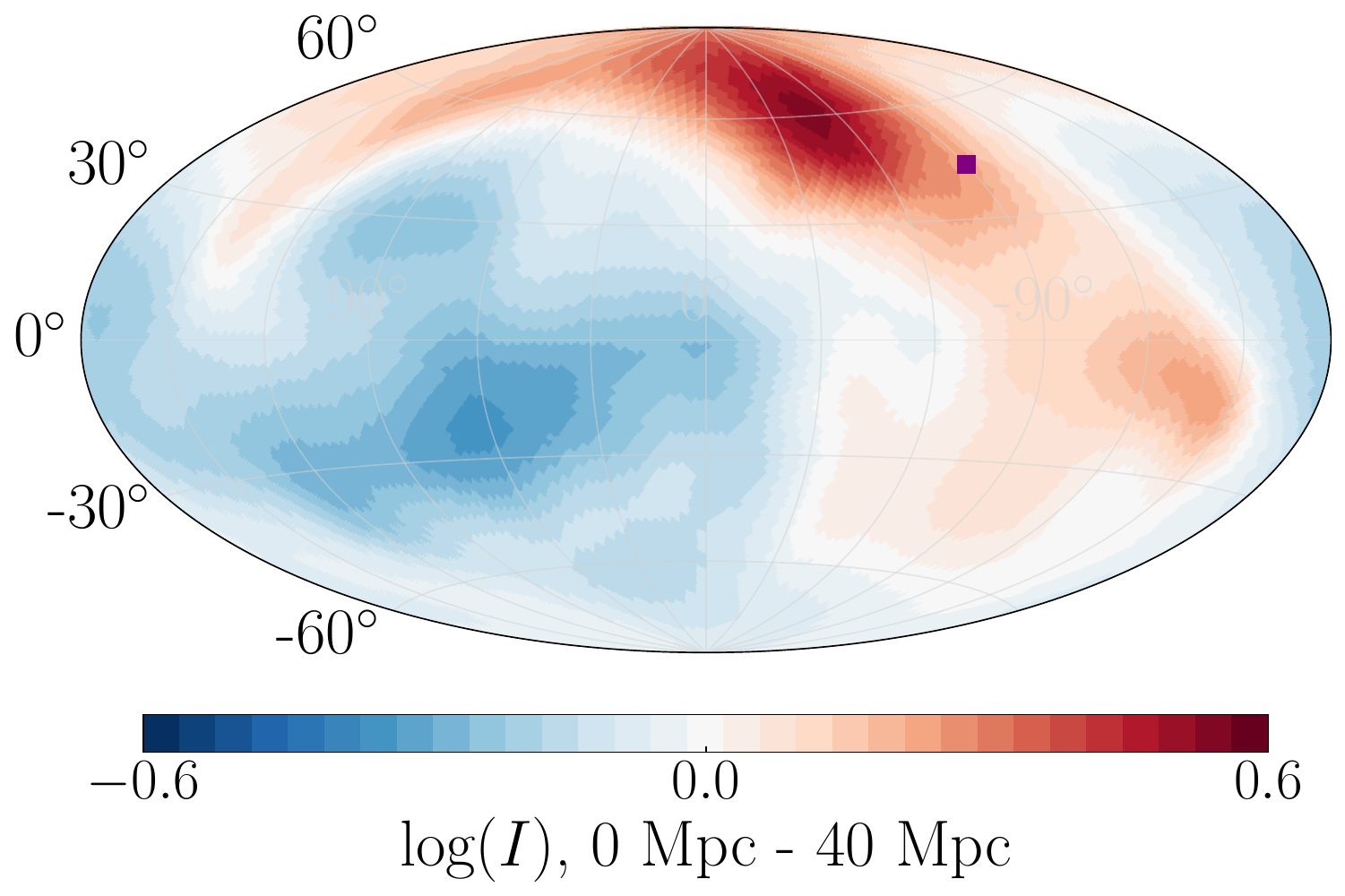}
        \vspace{\baselineskip}
        \includegraphics[width=0.49\textwidth]{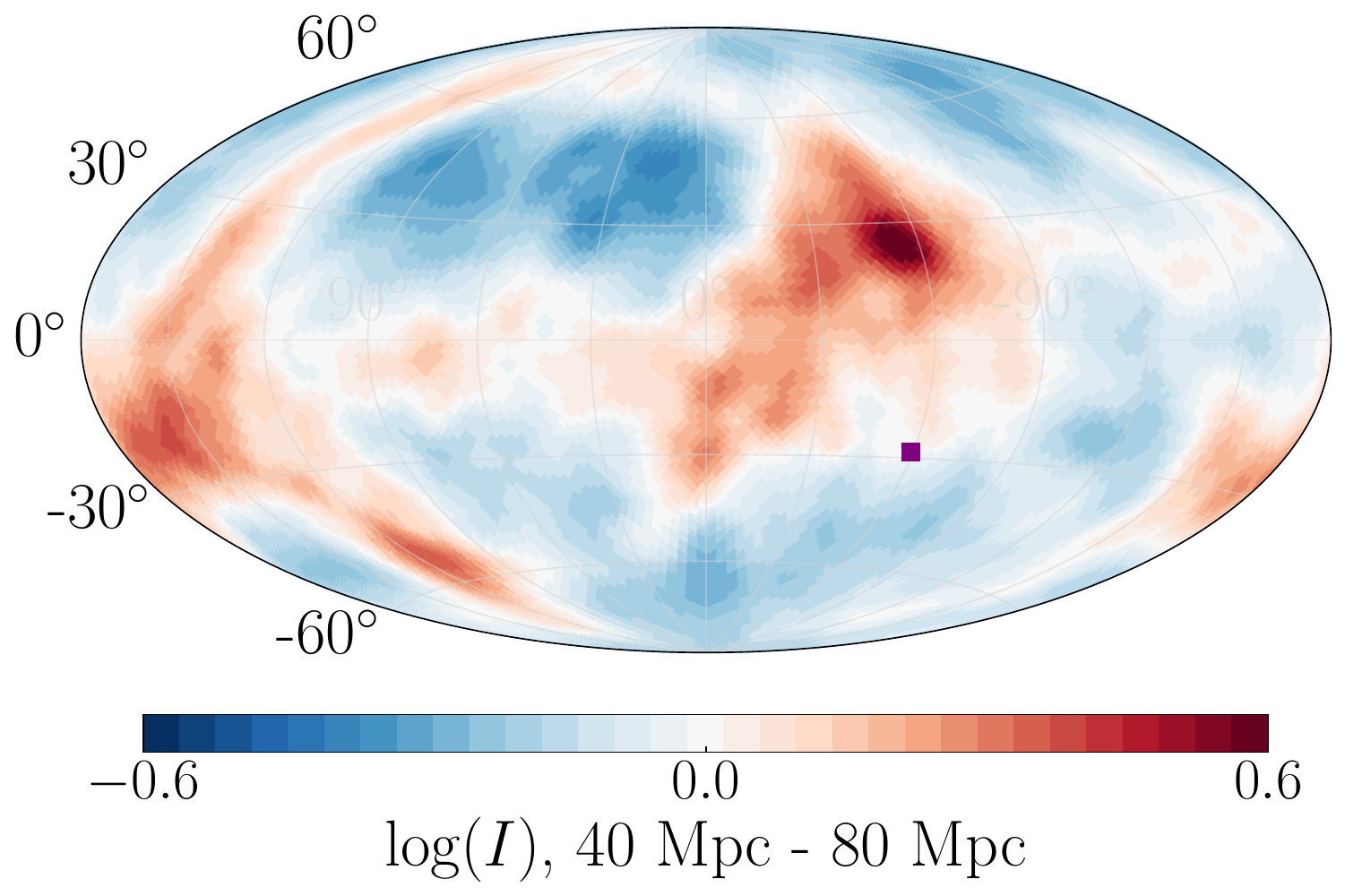}
        \includegraphics[width=0.49\textwidth]{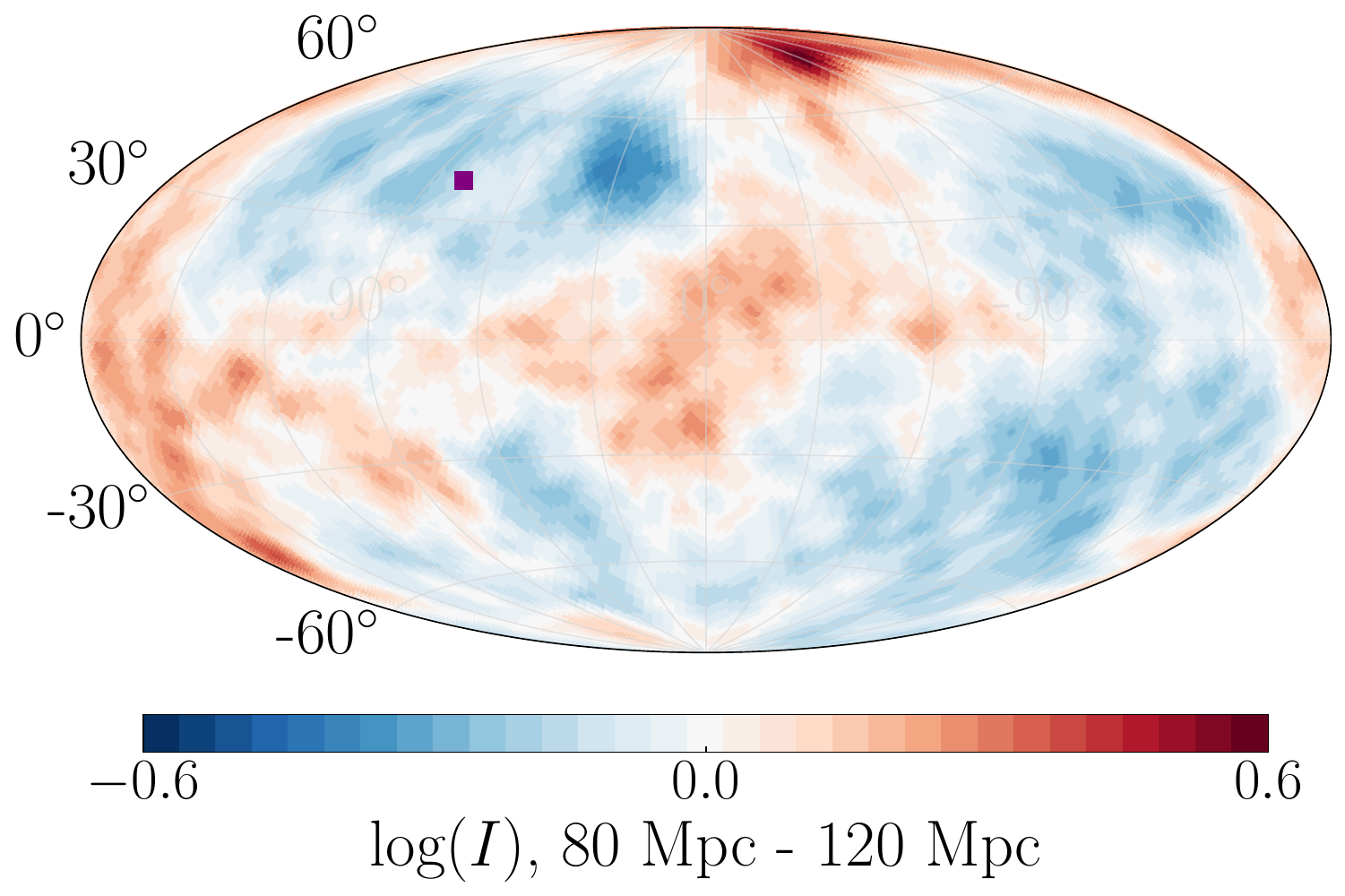}
        \includegraphics[width=0.49\textwidth]{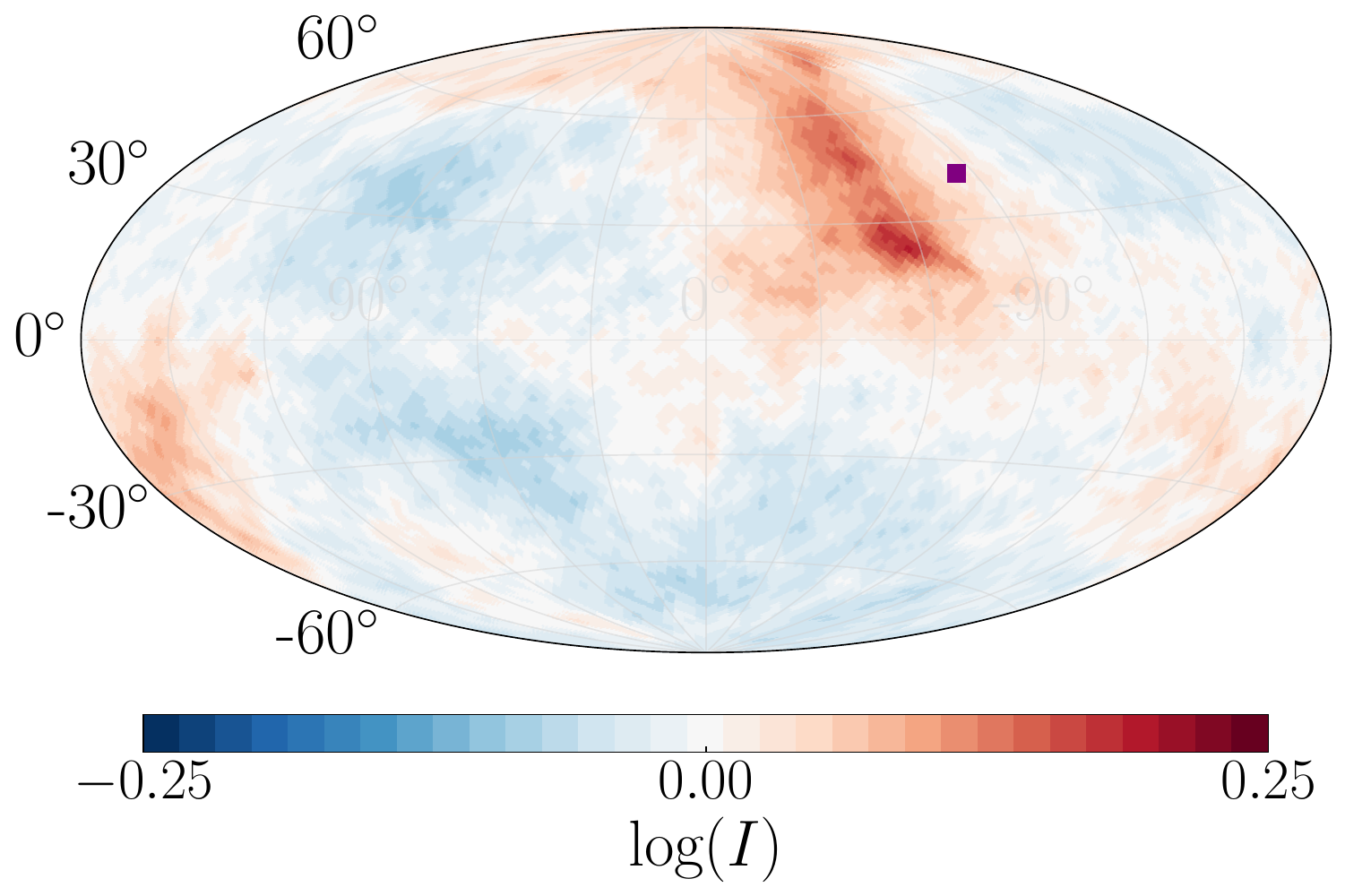}
        \caption{Logarithmic illumination maps for different distance bins and a cumulative distance bin (\textit{lower right}), for the best-fit model. The direction of the full-sky dipole component of the map is indicated with a purple marker.} \label{fig:illum}
    \end{minipage}
\end{figure}

We model the UHECR sources following the matter density of CosmicFlows 2 between 0 and 360\,Mpc. Beyond that up to 5000\,Mpc, we use a linear extrapolation of the density and assume a homogeneous distribution of sources. The source density as a function of distance is shown in Fig.~\ref{fig:propagation}. In analogy to previous works by Auger~\cite{Auger_CF_2017, Auger_CF_2023, Auger_CFAD_2023}, we assume an injection following the Peters cycle leading to a broken-exponential with spectral index $\gamma$ and a cutoff at the maximum rigidity $R_\mathrm{cut}$. We inject and track five representative elements: H, He, N, Si, and Fe.
Following the results from~\cite{Ding_2021}, we assume a negligible extragalactic magnetic field.
For the propagation, we use a database of 1-dimensional CRPropa3~\cite{CRPropa} simulations based on the Gilmore~\cite{Gilmore} and TALYS~\cite{TALYS} models. The interactions with background photon fields lead to a smaller propagation horizon for nuclei with larger energies, as displayed in Fig.~\ref{fig:propagation}. At the edge of the galaxy, we calculate so-called \textit{illumination maps} $I$ (using healpy pixelization~\cite{healpy}), which visualize the illumination of the Galaxy by the local anisotropic source distribution shaped by propagation effects. In Fig.~\ref{fig:illum}, the best-fit illumination maps for different distance bins are displayed. Here, it is visible that the local matter distribution is peaked in the Galactic north where for example the Virgo cluster resides.

The local arrival direction distribution can be calculated from the illumination maps by using a \textit{lens} of the GMF. Here, we use the same high-resolution backtracking simulations as~\cite{Ding_2021} of the JF12 model~\cite{JF12a, JF12b} with a coherence length of 30\,pc. After considering the effect of exposure of the Pierre Auger Observatory, we acquire the modeled arrival directions in energy bins $(8-16)$\,EeV, $(16-32)$\,EeV, and \textgreater32\,EeV, for which we calculate the dipole directions $\Vec{d}_E=(d_x, d_y, d_z)_E$ considering the effect of the limited exposure (see K-inverse method in~\cite{Ding_ICRC_2019}).
The dipole components $d_{{i\in(x,y,z)}_E}$ for each of the three energy bins are then compared to the latest official measurements by Auger~\cite{Almeida_dipole_2021} using a Gaussian likelihood function. Additionally, as in~\cite{Auger_CF_2023}, we compare the modeled energy spectrum to the unfolded spectrum from~\cite{Auger_E_2020} via a Poissonian likelihood in energy bins of width $\log_{10}(E)=0.1$ above $8\,\mathrm{EeV} \approx 10^{18.9}\,\mathrm{eV}$. To ensure that the model composition agrees with the measurements at Earth, we calculate the expected maximum shower depths \xmax from the arriving masses and energies using the EPOS-LHC hadronic interaction model~\cite{EPOS}. Similar to~\cite{Auger_CFAD_2023}, we consider a possible shift of the \xmax scale following the experimental systematic uncertainty as a nuisance parameter in the model. The modeled shower depths are then compared to the data~\cite{Yushkov_xmax_2019} via a Multinomial likelihood as in~\cite{Auger_CF_2017, Auger_CF_2023, Auger_CFAD_2023}.

\section{Results}
The measured spectrum and composition (not shown) are well described by the model. The best-fit injection is in agreement with recent results using a homogeneous source model~\cite{Auger_CF_2023, Auger_CFAD_2023}, with a hard spectral index $\gamma=1.2$, small maximum rigidity $R_\mathrm{cut}=10^{18.2}$\,eV, a best-fit shift of the \xmax scale of $-0.9\sigma$, and a nitrogen-dominated composition.
\begin{figure}[ht]
    \centering
    \begin{minipage}[b]{0.44\linewidth}
        \centering
        \includegraphics[width=\textwidth]{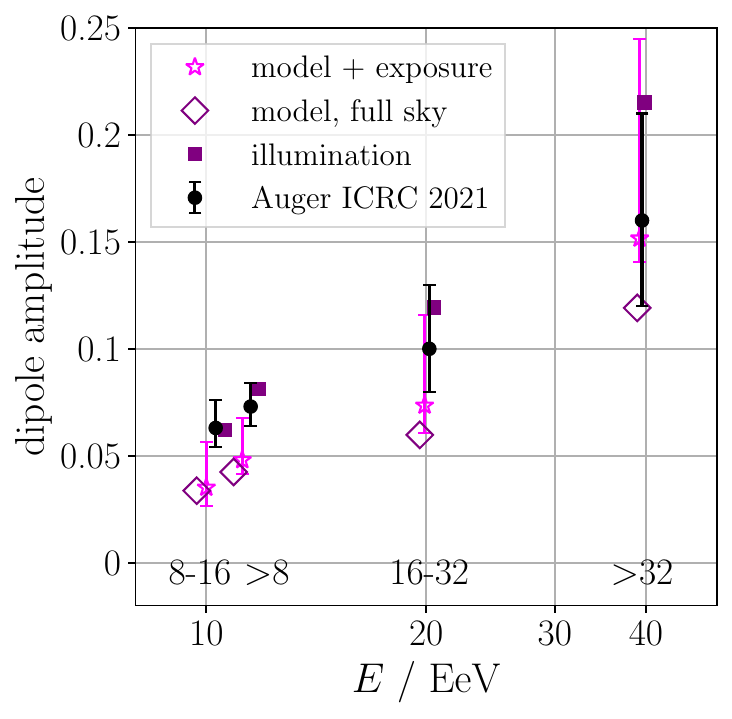}
        \caption{Best-fit dipole amplitude: black are data from~\cite{Almeida_dipole_2021} ($d$ in tab. 1); purple filled squares the illumination (Fig.~\ref{fig:illum}); purple unfilled squares the whole sky; pink stars include exposure effect, with 68\% C.L. stat. uncertainty. The markers are offset on the x-axis for better visibility.} \label{fig:dipole_amp}
    \end{minipage}
    \hfill
    \begin{minipage}[b]{0.54\linewidth}
        \centering
        \includegraphics[width=0.49\textwidth]{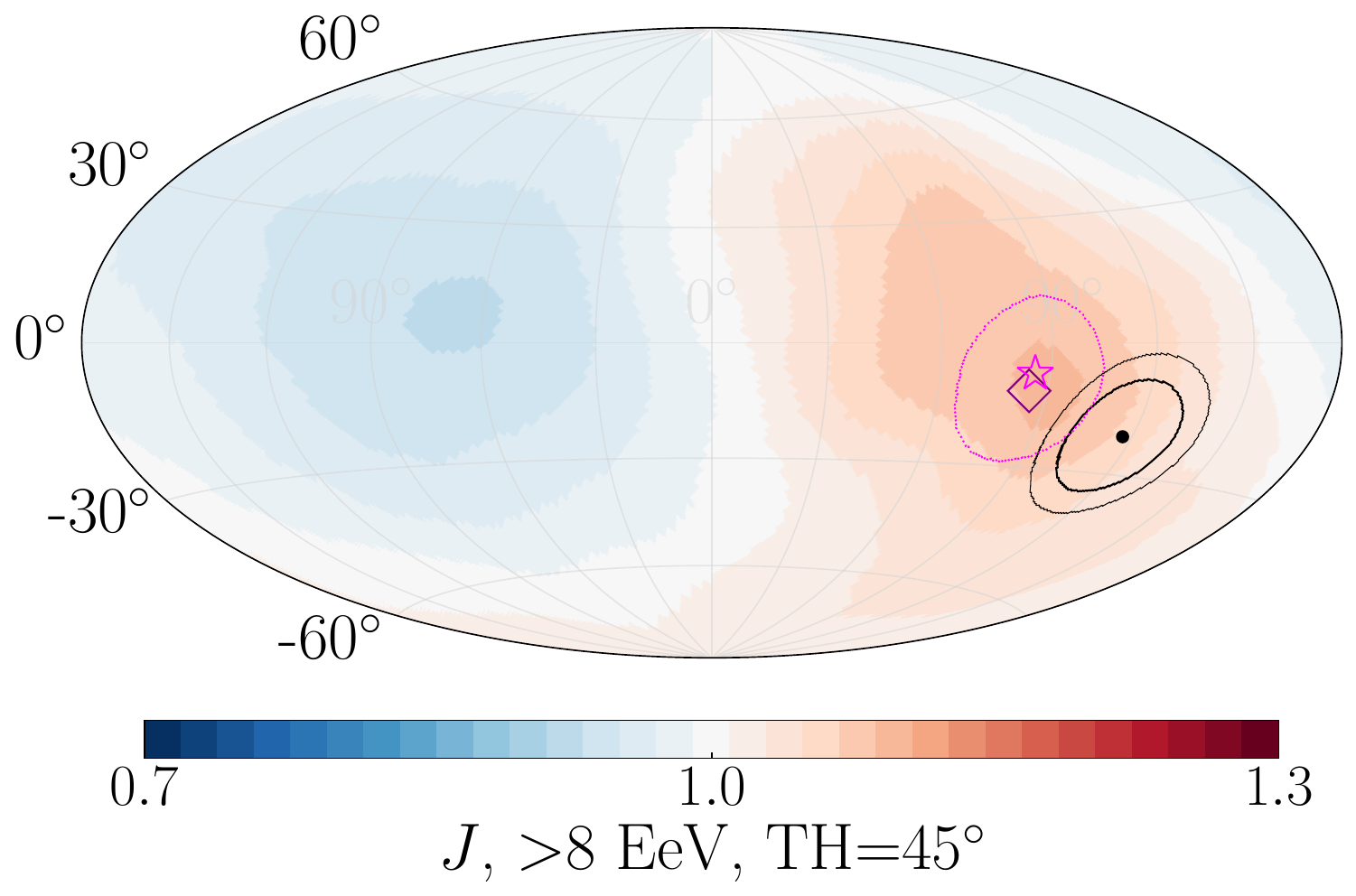}
        \vspace{\baselineskip}
        \includegraphics[width=0.49\textwidth]{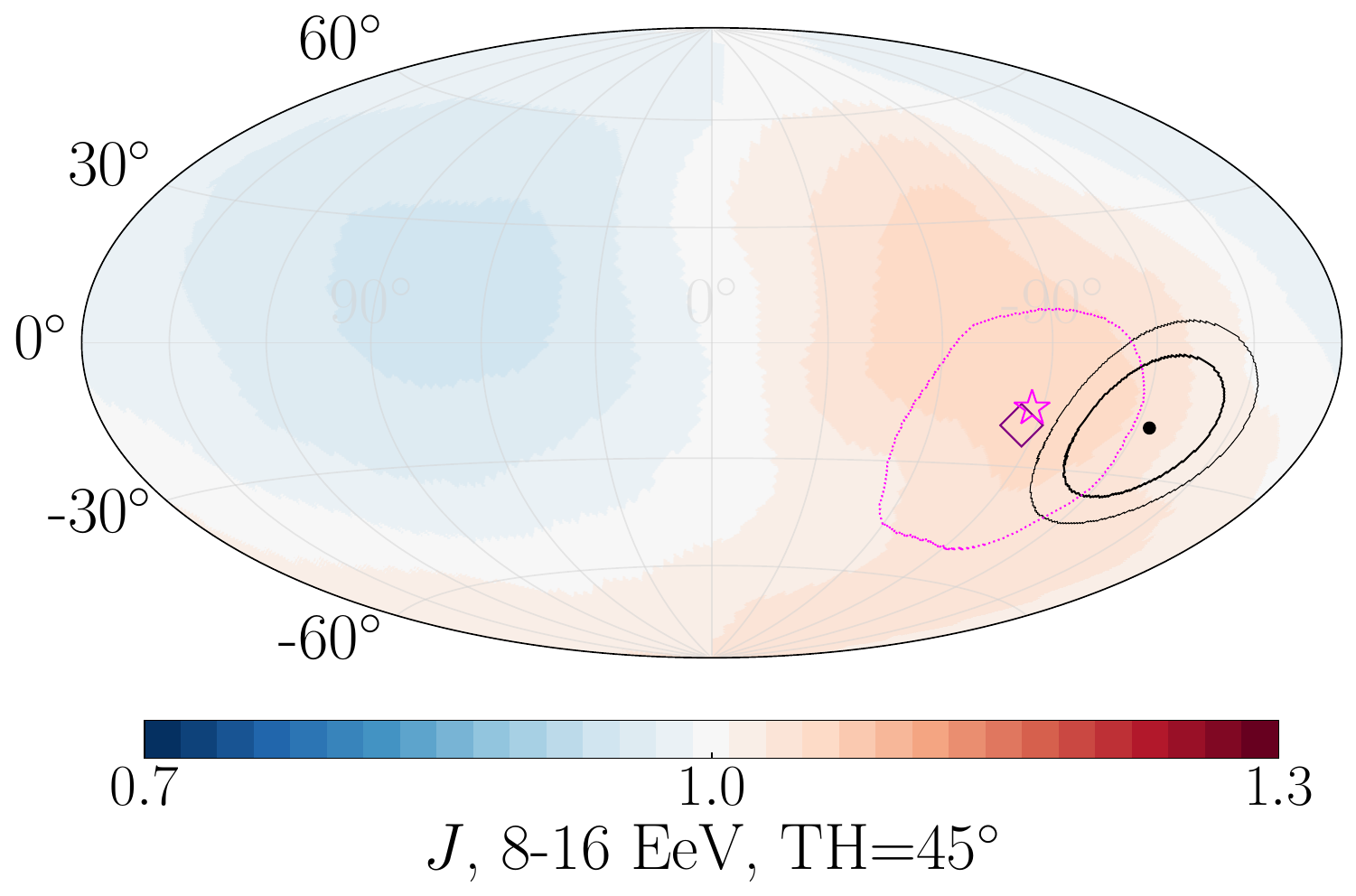}
        \vspace{\baselineskip}
        \includegraphics[width=0.49\textwidth]{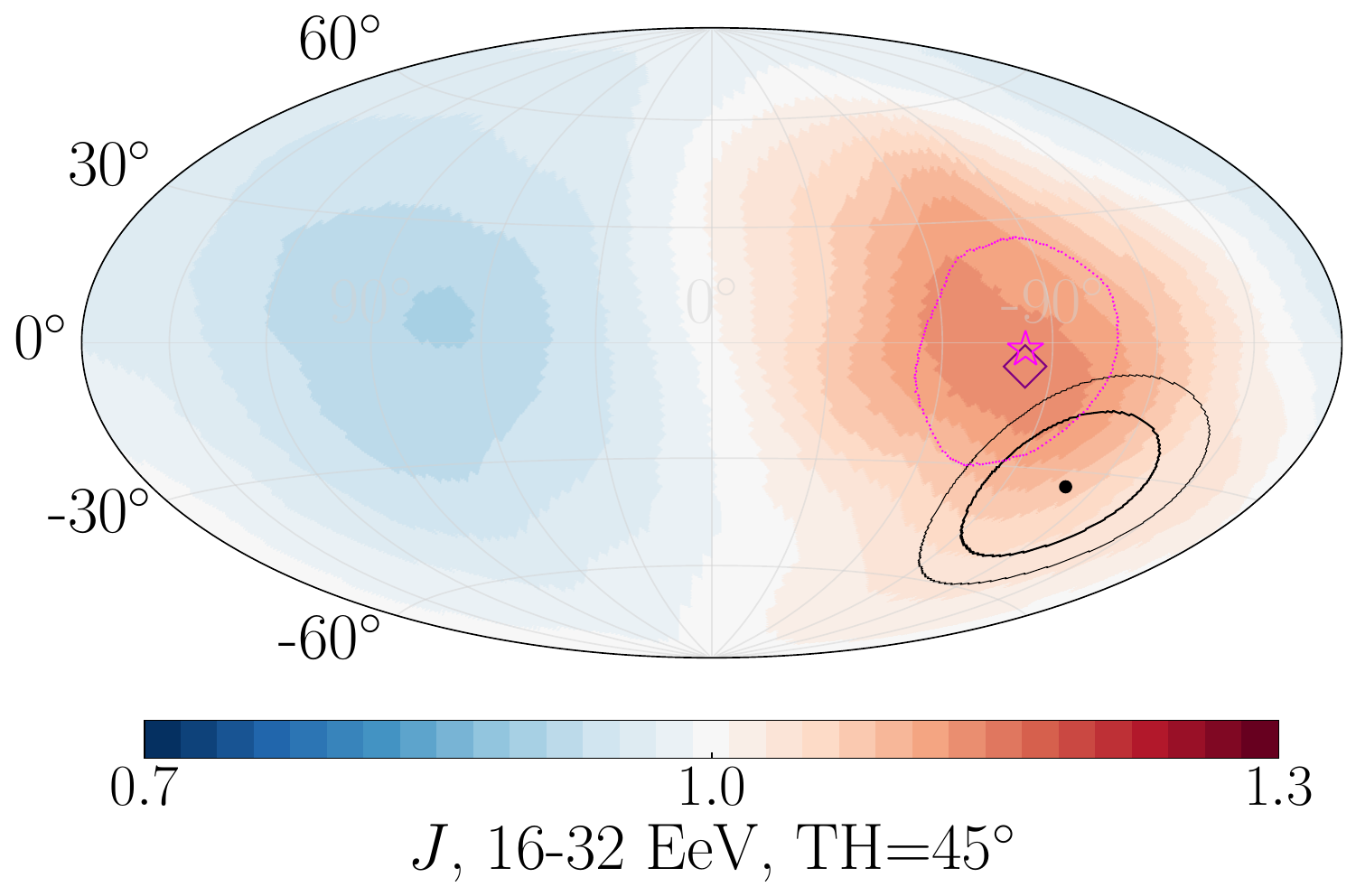}
        \vspace{\baselineskip}
        \includegraphics[width=0.49\textwidth]{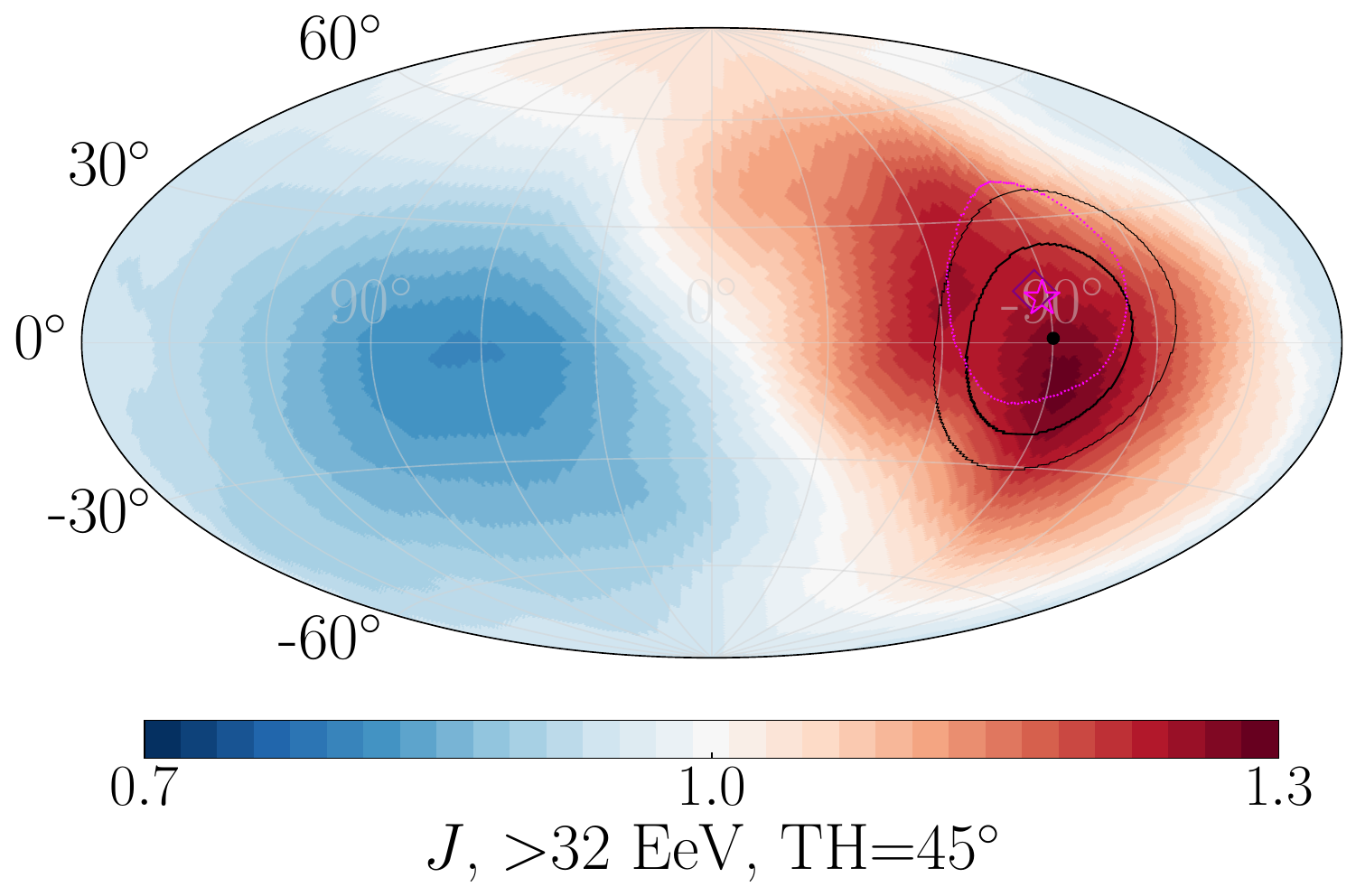}
    \caption{Best-fit modeled directional flux $J$; same markers indicating the dipole directions as in Fig.~\ref{fig:dipole_amp}; 68\% C.L. stat. uncertainty of the best-fit dipole direction in pink; $45^\circ$ tophat smoothing. The $68\%$ and $90\%$ contours of the dipole measured by Auger~\cite{Almeida_dipole_2021} are indicated by black ellipses.} \label{fig:dipole_direc}
    \end{minipage}
\end{figure}

The fitted dipole amplitude and direction are shown in Fig.~\ref{fig:dipole_amp} and Fig.~\ref{fig:dipole_direc}. In both figures, the best-fit result including the exposure effect (magenta) is depicted with its 68\% C.L. statistical uncertainty from the variations due to the limited number of events (set to 44000 following~\cite{Almeida_dipole_2021}).
Both best-fit dipole amplitude and direction are very close to the measurements for the highest energy bin \textgreater32\,EeV. For smaller energies, however, the best-fit dipole amplitude is slightly too small, and the direction is somewhat outside the 2$\sigma$ region. We expect that both of these issues can be resolved with an updated model of the GMF~\cite{UF23} with reduced turbulent field strength and slightly modified directional deflections. Nevertheless, it has to be noted that
\begin{wrapfigure}[17]{r}{7.5cm}
\includegraphics[width=0.43\textwidth]{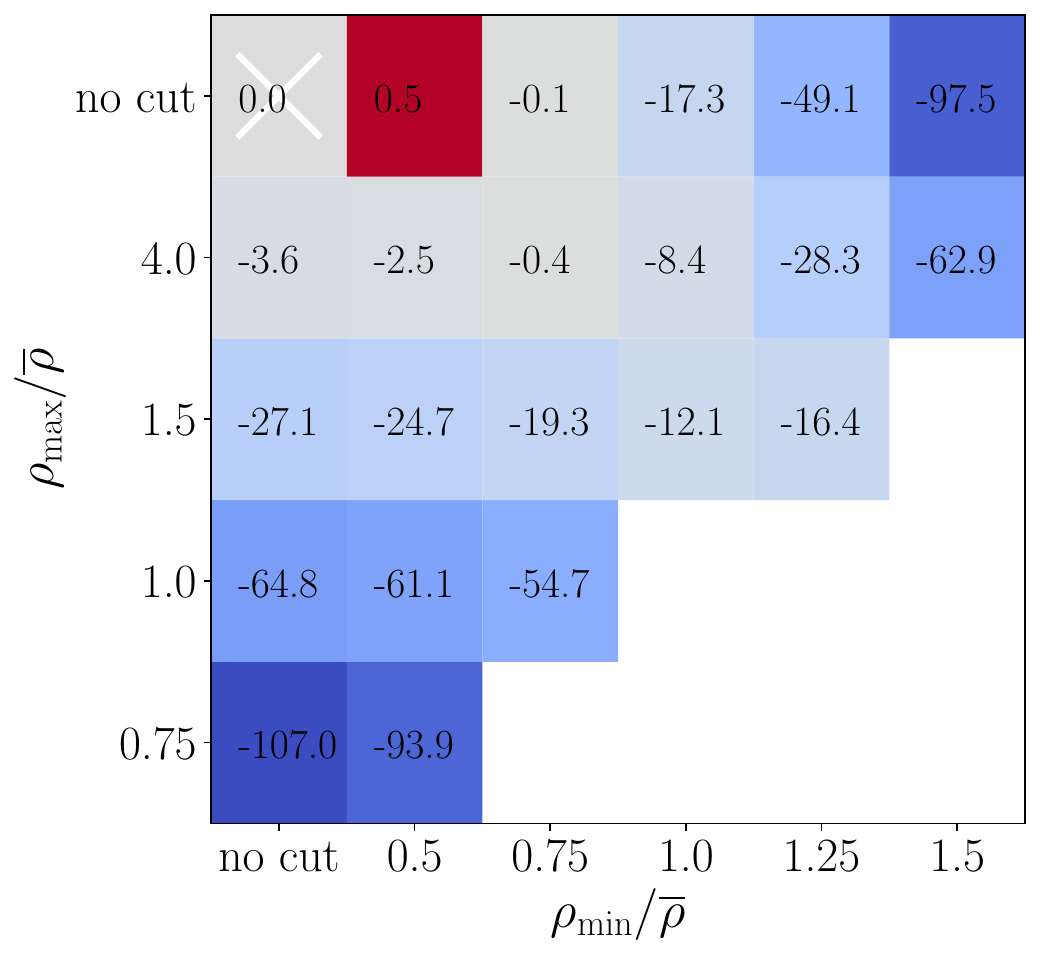}
\caption{Influence of a bias: regions with densities below $\rho_\mathrm{min}$ or above $\rho_\mathrm{max}$ are excluded from the UHECR injection during the fit. Color / numbers indicate the likelihood ratio ($\log \mathcal{L} - \log \mathcal{L}_x$) compared to the model with no cuts (marked with white \textit{x}).}
\label{fig:bias}
\end{wrapfigure}
 the general effect of an increasing amplitude, and a relatively steady direction that moves towards the Galactic north at the highest energies can already be described well within the expected uncertainties when using the current JF12 field.


\paragraph{Effect of ``bias'':} We investigate the effect of a possible \textit{bias} in the relation between UHECR sources and the underlying LSS distribution by excluding regions of the source distribution in the fit if they either exceed a maximum density $\rho_\mathrm{max}$, or are lower than a minimum density $\rho_\mathrm{min}$ (where the mean density is denoted as $\overline{\rho}$). The results are shown in Fig.~\ref{fig:bias}.
It is visible that the likelihood decreases for almost any value of the cuts. Without the high-density regions, the sky becomes more isotropic. Then, the dipole direction is not represented well anymore, especially as even the cut $\rho_\mathrm{max}=4\overline{\rho}$ removes parts of the nearby Virgo cluster in the Galactic north which is important to reproduce the observed dipole. Removing low- and intermediate-density regions with $\rho\lesssim\overline{\rho}$ leads to too large anisotropies, overshooting the dipole amplitude. Only a cut removing the very low-density regions below $0.5\cdot\overline{\rho}$ leads to a slight, not-significant improvement of the likelihood. This study suggests that UHECR sources seem to reside in high- and medium-density regions, and no definite conclusion can be drawn regarding low-density regions.

\begin{figure}[ht]
\includegraphics[width=0.33\textwidth]{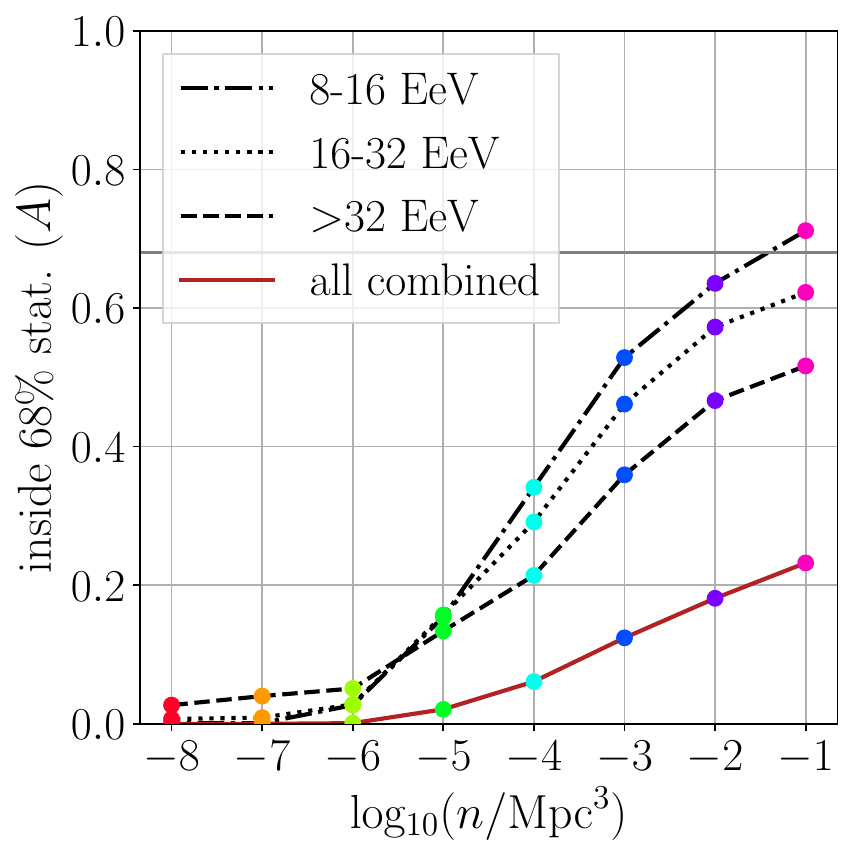}
\includegraphics[width=0.33\textwidth]{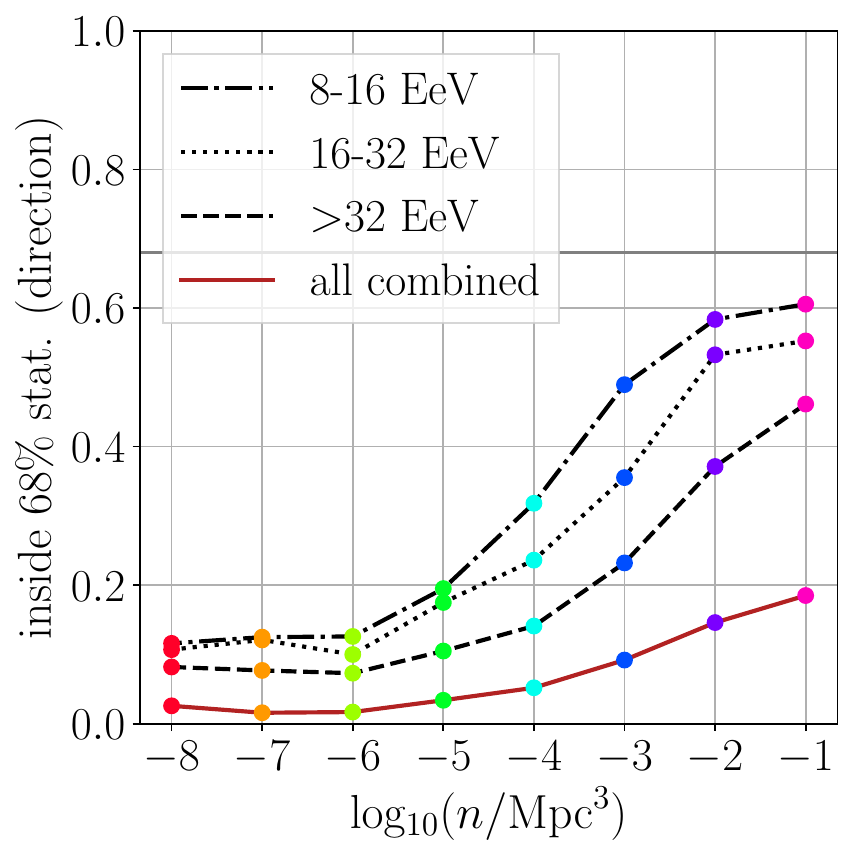}
\includegraphics[width=0.33\textwidth]{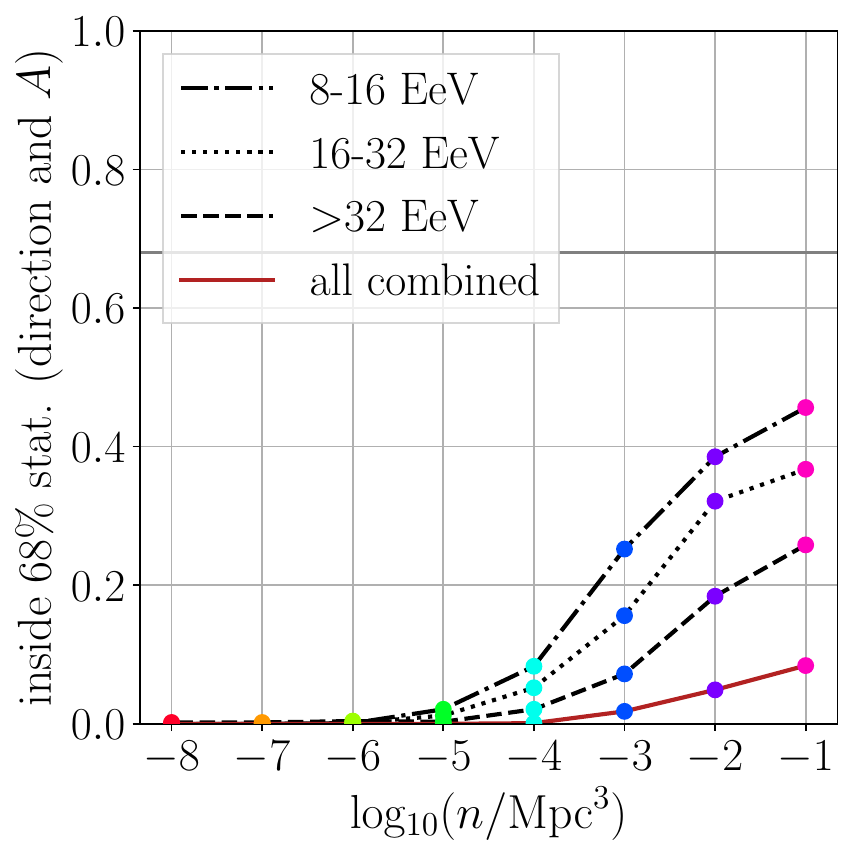}
\caption{Fraction of realizations where dipole amplitude (\textit{left}) / direction (\textit{middle}) / both at the same time (\textit{right}) are within the 68\% uncertainty of the continuous model depending on the source number density $n$.}
\label{fig:inside_fraction}
\end{figure}

\paragraph{Limits on the source number density:}
Building on the best-fit model presented in the previous sections (in the following called \textit{continuous model}), we will now place constraints on the source number density. For that, we draw explicit source catalogs 1000 times from the complete source distribution used above. Then we draw 44000 events following the event statistic from~\cite{Almeida_dipole_2021} and investigate the influence on the dipole direction and amplitude, as well as the whole power spectrum. In Fig.~\ref{fig:inside_fraction} the fraction of realizations where the dipole amplitude and/or direction is within the $68\%$ statistical uncertainty\footnote{In future works with updated GMF so the continuous model gives an accurate description of the dipole amplitude and direction, the fraction inside the \textit{measured} 68\% region will be used instead of inside the \textit{modeled} statistical uncertainty.} of the continuous model (from the measured number of events, as shown in Fig.~\ref{fig:dipole_amp} and Fig.~\ref{fig:dipole_direc}) is depicted. It is evident that for densities $n\lesssim 10^{-4}$ / Mpc$^3$ the amplitudes become too large, and the dipole directions too random, so that only 1 out of the 1000 realizations reproduces both measures at the same time for all energy bins at $n=10^{-4}$ / Mpc$^3$.

\begin{figure}[ht]
\includegraphics[width=0.49\textwidth]{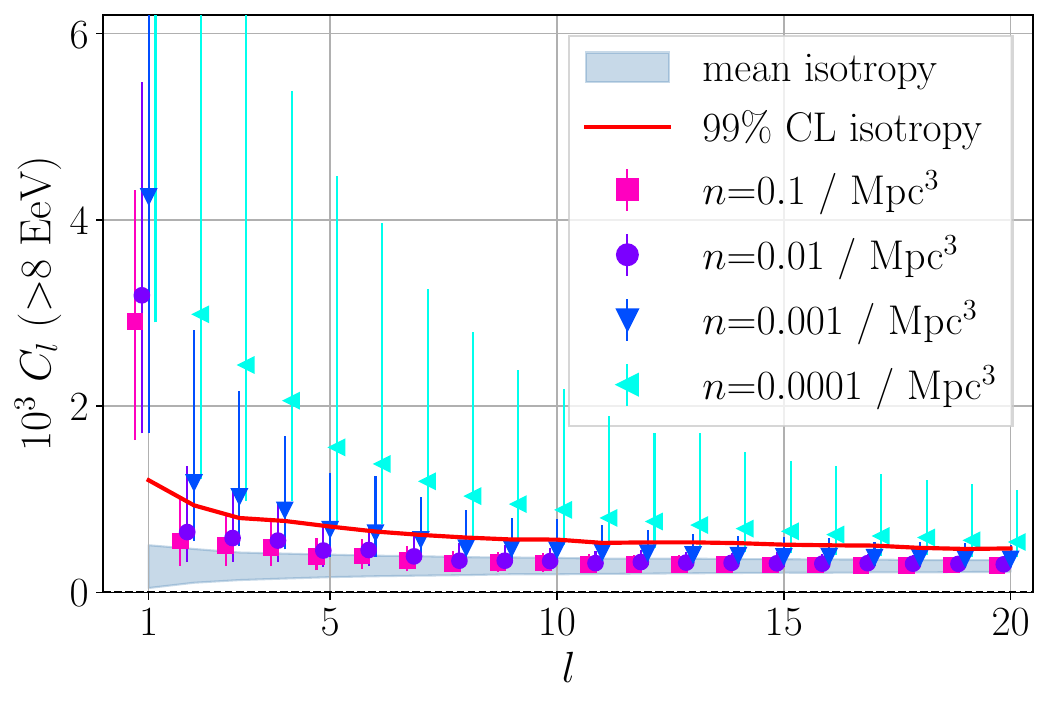}
\includegraphics[width=0.49\textwidth]{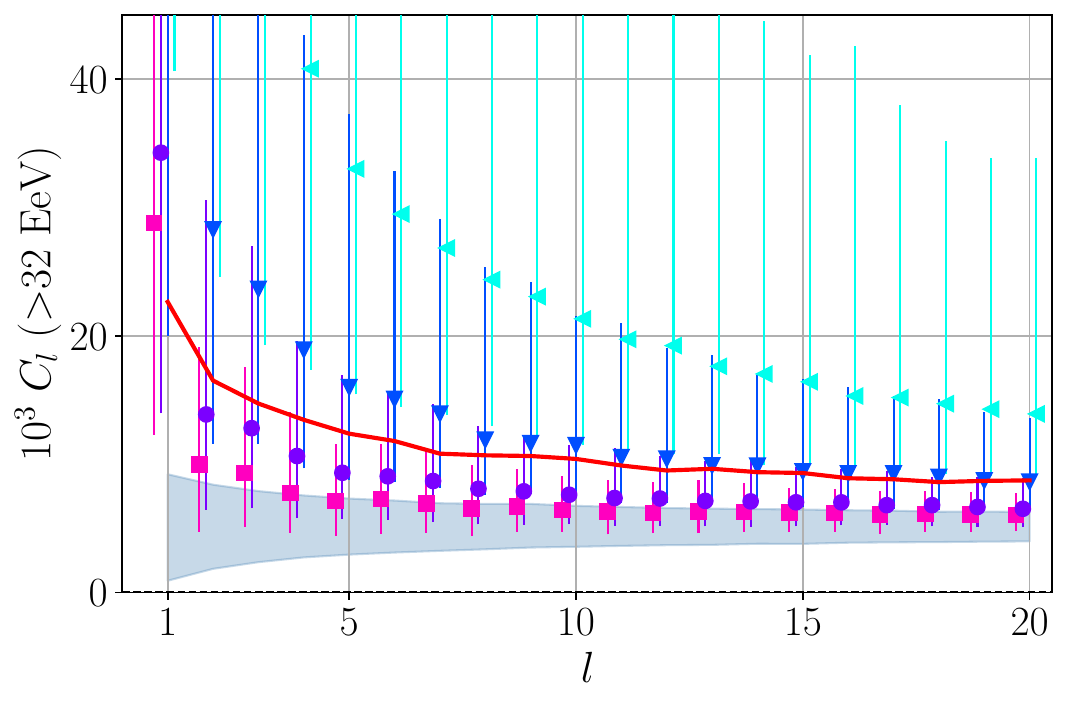}
\caption{Angular power spectrum for different source number densities, for $E>8$ and 32 EeV datasets (left and right panels). The markers show the median value and the ``error bars" indicate the range encompasing $68\%$ of the cases including statistical fluctuations in source and cosmic ray samples. The 99\% expectation from isotropic arrival directions is shown as a red line, mean $\pm$ standard deviation as a grey band. Note that for Auger data~\cite{Almeida_dipole_2021} all $C_{l>1}$ are within isotropic expectations.}
\label{fig:power_spectrum}
\end{figure}

In addition to the dipole amplitude and direction, we also study the effect of the source density on the angular power spectrum. This is displayed in Fig.~\ref{fig:power_spectrum} for two cumulative energy bins. Note that for densities $\gtrsim 10^{-2}$ / Mpc$^3$ (and also the continuous model) most higher moments $C_{l>1}$ are within $99\%$ expectations from isotropy as also seen in the data~\cite{Almeida_dipole_2021}.
The number of random realizations with a dipole moment larger than 5\% for the cumulative energy bin >8\,EeV, when at the same time requesting all higher multipoles $C_{l>1}$ to be within 99\% of the isotropic expectation, is 134/1000 for a number density of $n=10^{-1}$ / Mpc$^3$,  121/1000 for $n=10^{-2}$ / Mpc$^3$ and  12/1000 for $n=10^{-3}$ / Mpc$^3$.  For $n\leq10^{-4}$ / Mpc$^3$, no realizations fulfill both criteria.

\begin{figure}[ht]
\begin{minipage}[b]{0.48\linewidth}
\includegraphics[width=\textwidth, align=c]{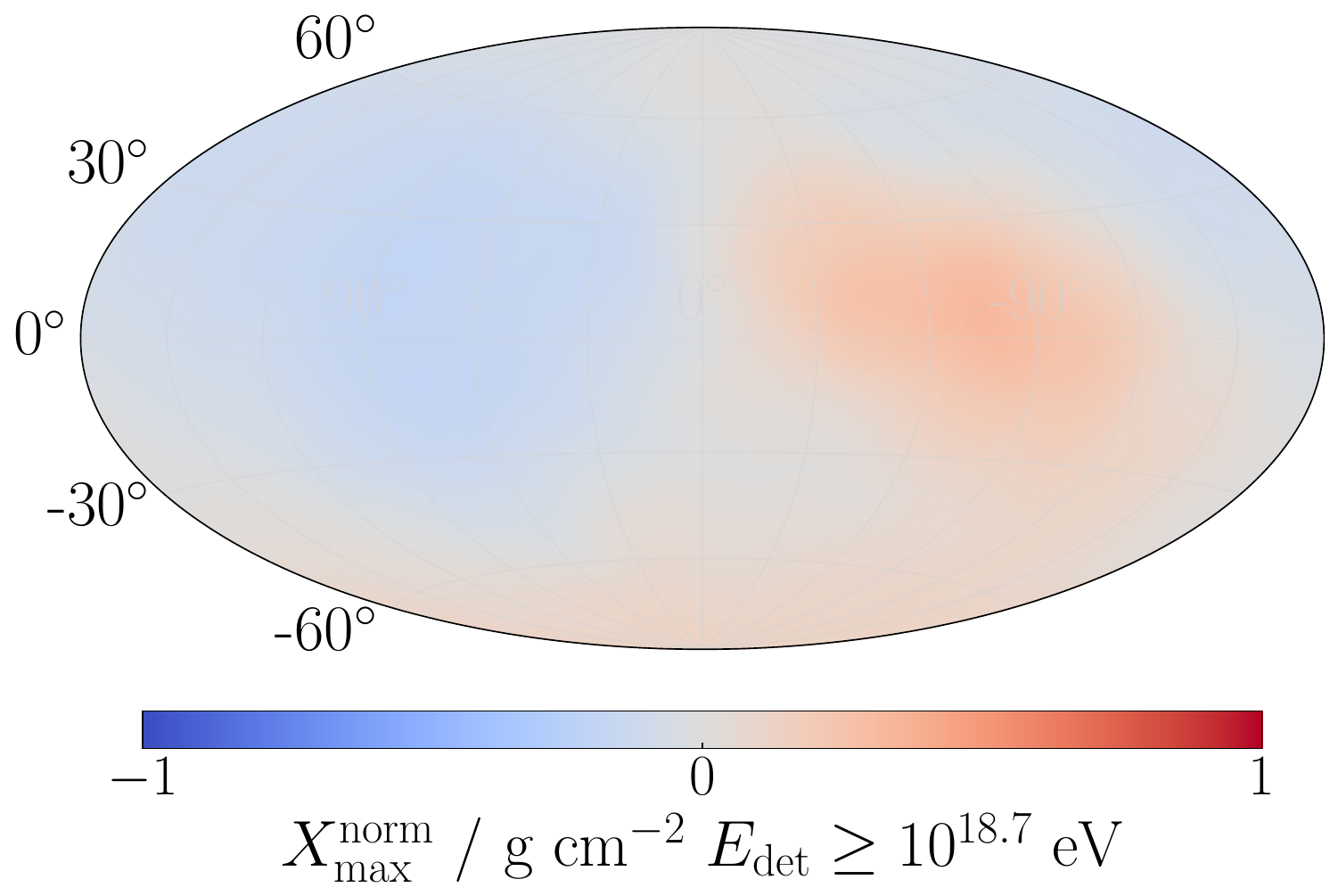}
\caption{Normalized (corrected for the energy evolution) maximum shower depth $\xmax^\mathrm{norm}$ for the continuous best-fit model with a $30^\circ$ tophat smoothing; in analogy to~\cite{Mayotte_ICRC_2021}.} \label{fig:heaviness_map}
\end{minipage}
\hfill
\begin{minipage}[b]{0.49\linewidth}
\includegraphics[width=\textwidth, align=c]{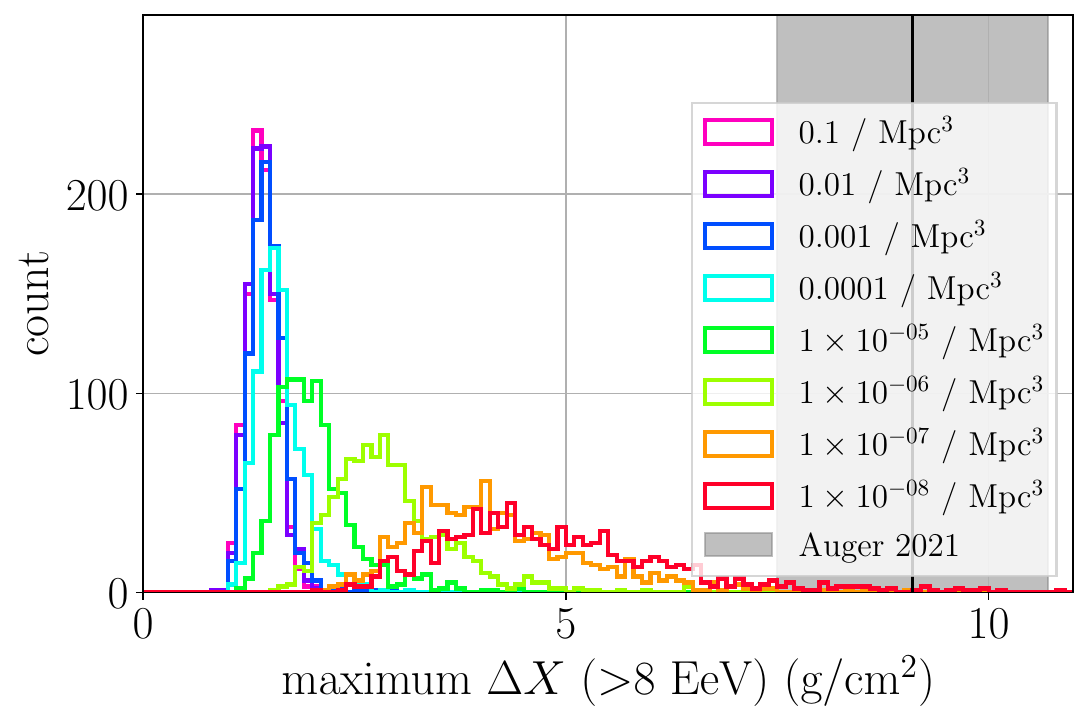}
\caption{Histogram of the maximum difference between smallest and largest $\xmax^\mathrm{norm}$ in $30^\circ$ tophat smoothed maps, depending on the source number density, assuming 4000 events.} \label{fig:heaviness_hist}
\end{minipage}
\end{figure}

\paragraph{Arrival-direction dependent composition anisotropies:}
Following the indication for a composition anisotropy observed in the Auger data~\cite{Mayotte_ICRC_2021} (note updated results~\cite{Mayotte_ICRC_2023}), we also calculate the normalized shower depth (corrected for the energy evolution effect) as a function of the direction, displayed in Fig.~\ref{fig:heaviness_map}. For the continuous model (\textit{left}), the composition does not vary greatly over the sky, with a maximum difference of $\Delta X \lesssim2$ g/cm$^2$. The lighter part (red) is correlated with the flux overdensity seen in Fig.~\ref{fig:dipole_direc}, which is expected since the lighter (low Z) component diffuses less in the GMF. The effect of the source number density on the composition anisotropy is visualized in Fig.~\ref{fig:heaviness_hist}, where the maximum difference between the smallest and largest $\xmax^\mathrm{norm}$ in $30^\circ$ tophat smoothed maps (like Fig.~\ref{fig:heaviness_map}) of the 1000 randomly drawn realizations is depicted. For the maps we use a smaller event statistic of 4000 as~\cite{Mayotte_ICRC_2021} is based on data by the fluorescence detector with limited duty cycle. It is visible how only very small densities, $n\lesssim10^{-6}$ g/cm$^2$, can lead to differences between on- and off-plane regions approaching the value of $9.8$ g/cm$^2$ reported in~\cite{Mayotte_ICRC_2021} . As described above, such small densities lead to very large flux anisotropies not in agreement with Auger data.

\begin{figure}[ht]
\includegraphics[width=0.33\textwidth]{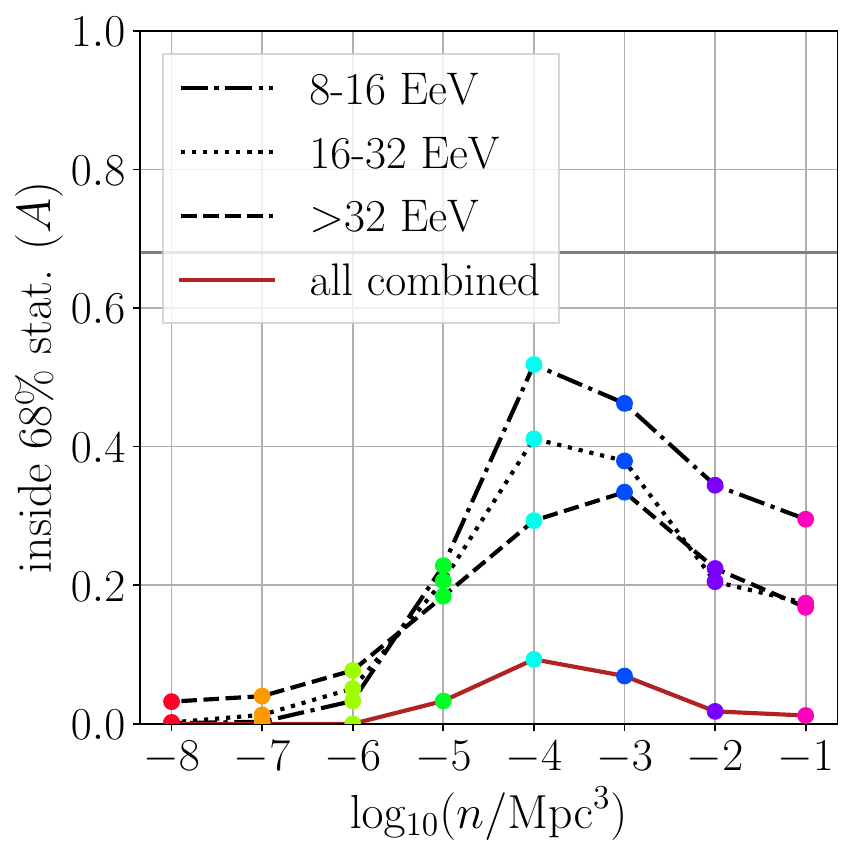}
\includegraphics[width=0.33\textwidth]{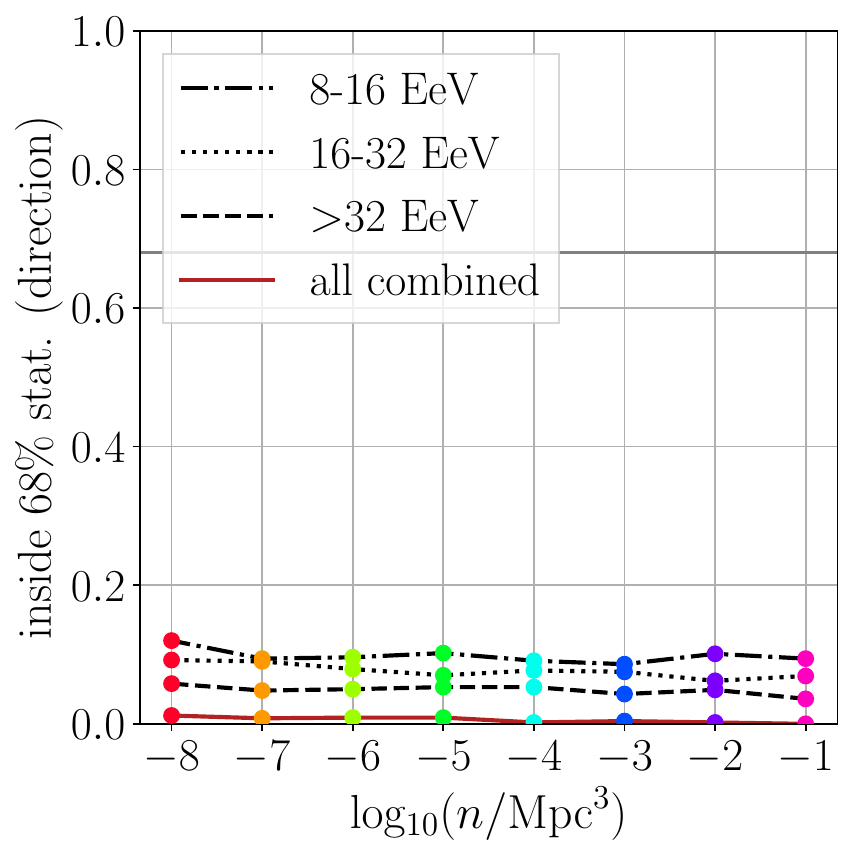}
\includegraphics[width=0.33\textwidth]{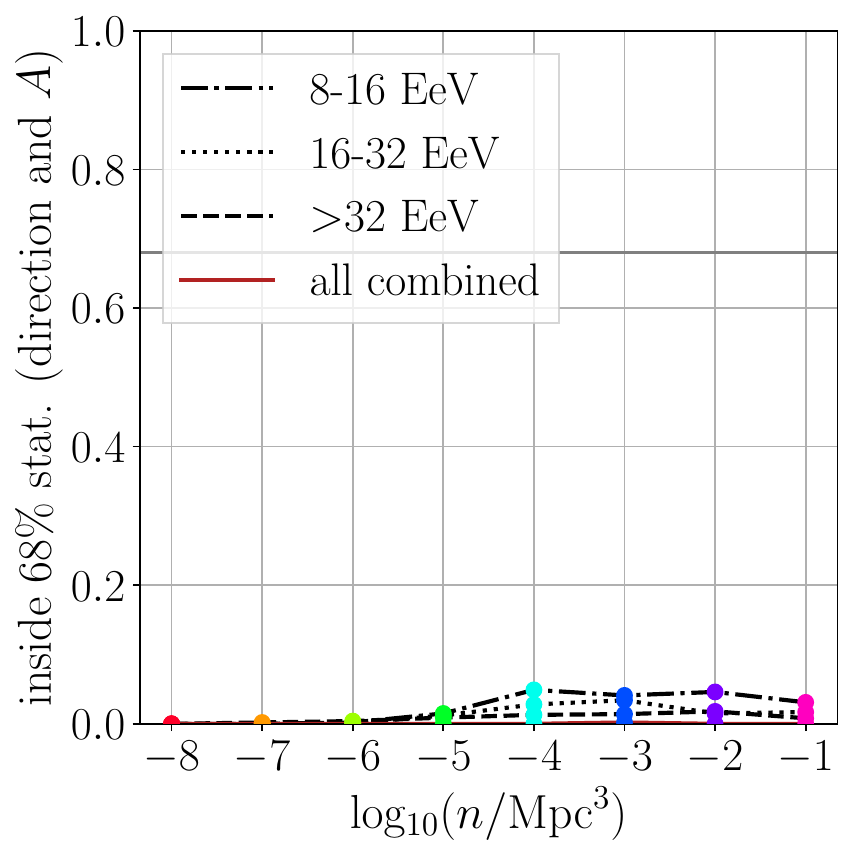}
\includegraphics[width=0.49\textwidth]{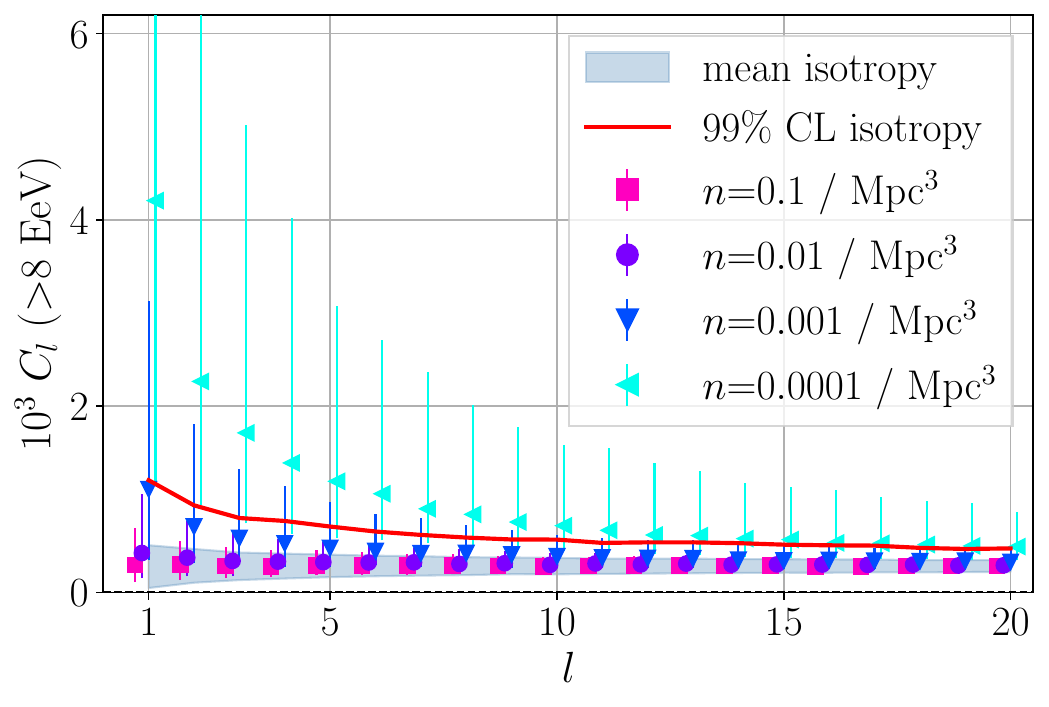}
\includegraphics[width=0.49\textwidth]{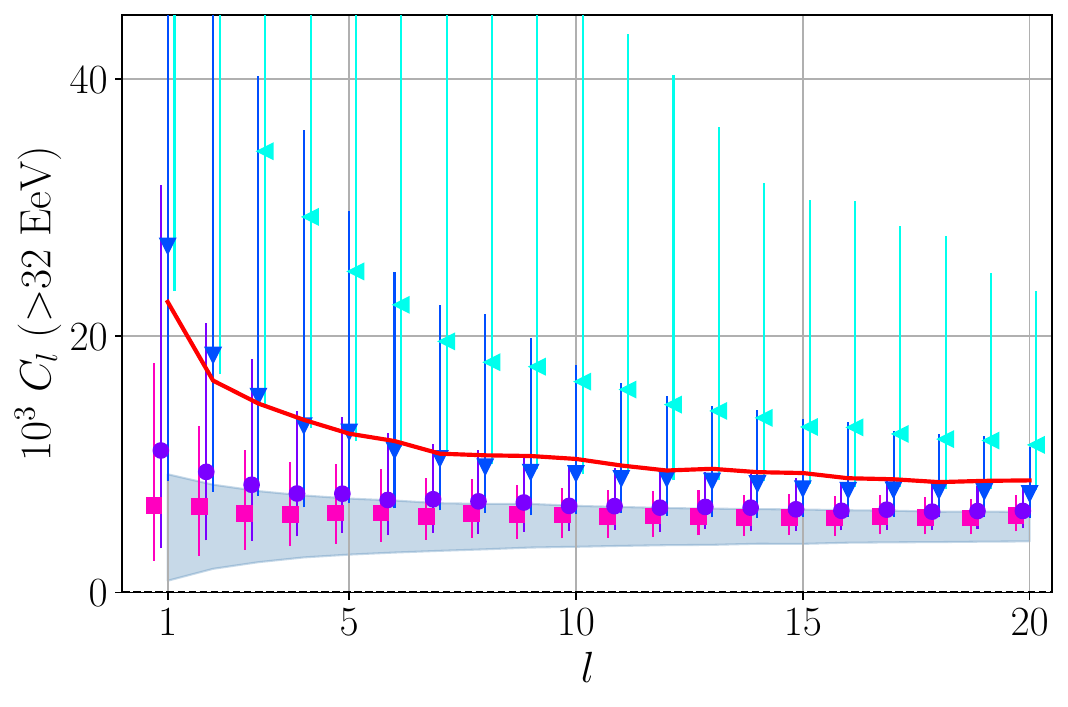}
\caption{Same as Figs.~\ref{fig:inside_fraction} and ~\ref{fig:power_spectrum}, but for a homogeneous source distribution not following the LSS.}
\label{fig:inside_fraction_iso}
\end{figure}

\paragraph{Isotropic source distribution:}
In~\cite{Auger_dipole_2018} it was suggested that homogeneously distributed sources not following the LSS, with a density of around $n\approx10^{-4}$ / Mpc$^3$, could also explain the measured dipole energy evolution. We investigate this by changing the source distribution to a fully homogeneous setup, and repeating all steps described above.  The composition and spectrum at the source are the same as for the LSS model. 
In the first panel of Fig.~\ref{fig:inside_fraction_iso} it is visible that a density of $n\approx10^{-4}$ / Mpc$^3$ indeed seems to fit best with the dipole amplitude. As the dipole direction is however fully random, the likelihood of reproducing both amplitude and direction simultaneously in multiple energy bins is small. 

The power spectra depicted in the lower panels of Fig.~\ref{fig:inside_fraction_iso} suggest that densities of order $n\approx10^{-4}$ / Mpc$^3$, as necessary to produce a sizeable dipole moment, at the same time lead to too large values for all larger multipole moments in tension with results by Auger~\cite{Almeida_dipole_2021}. The number of realizations with a dipole moment larger than 5\% in the cumulative energy bin >8\,EeV for a number density of $n=10^{-4}$ / Mpc$^3$ is 565/1000, but none of these exhibits all higher multipoles $C_{l>1}$ within 99\% of the isotropic expectation. Also, for no other tested source density is a realization found that fulfills both criteria at the same time, in the case of a homogeneous source distribution.

We have also investigated the possibility of arrival-direction-dependent composition anisotropies for the homogeneous source model, and have found that while the directions of possible anisotropies in the sky are more random in that case, the size of the composition anisotropy does not depend on the source model (homogeneous and LSS source distribution lead to similar values of $\Delta X$).


\section{Conclusion}
We have demonstrated that the observed dipole anisotropy in the UHECR arrival directions~\cite{Auger_dipole_2017} and its energy dependency~\cite{Auger_dipole_2018, Almeida_dipole_2021} can be well described by a model where the UHECR sources follow the Large Scale Structure of the nearby universe. For that, we have updated previous works by C. Ding, N. Globus, and G. Farrar~\cite{Ding_2021, Ding_ICRC_2019} with a self-consistent treatment of propagation effects and a simultaneous fit to the spectrum, composition, and dipole data of the Pierre Auger Collaboration. 

Going further, we have investigated the possibility of bias between UHECR source and LSS distributions, and found that UHECR sources reside in both high- and intermediate-density regions, with no definite conclusion regarding low-density regions. We have constrained the number density of UHECR sources to $n > 10^{-4}$ / Mpc$^3$ by investigating the effect of density variations on the dipole amplitude, direction, and the power spectrum. Additionally, we have demonstrated that while it is in principle possible to reproduce the dipole energy evolution with a homogeneous source distribution with $n\approx10^{-4}$ / Mpc$^3$, this scenario leads to too-large higher multipole moments, not in agreement with Auger data. Also, we have shown that a very small number density $n\lesssim10^{-6}$ / Mpc$^3$ is necessary to produce measurable arrival-direction dependent composition anisotropies over the sky, which at the same time would lead to substantial flux anisotropies not compatible with current measurements.

In the future, we plan to investigate the impact of updates of the LSS model~\cite{CF3}, variations of the injection spectrum parameterization and hadronic interaction models, as well as different extragalactic magnetic field models. Additionally, we plan to constrain possible variations of the Galactic magnetic field~\cite{UF23} with our model.

%
%
%


\begin{thebibliography}{99}
\footnotesize
\setlength{\itemsep}{0.1pt}

\bibitem{Auger_dipole_2017}
A. Aab et al. (The Pierre Auger Collaboration), Science \textbf{357} 6357 (2017)

\bibitem{Auger_dipole_2018}
A. Aab et al. (The Pierre Auger Collaboration), ApJ \textbf{868} 1 (2018)

\bibitem{Ding_2021}
C. Ding, N. Globus and G. Farrar, ApJL \textbf{913} L13 (2021)

\bibitem{Allard_dipole_2021}
D. Allard, J. Aublin, B. Baret and E. Parizot, A\&A \textbf{664} A120 (2022)

\bibitem{Hoffman_CF}
Y. Hoffman, E. Carlesi,  D.Pomarede et al. Nat. Astron., \textbf{2}, 680 2018

\bibitem{Auger_CF_2017}
A. Aab et al. (The Pierre Auger Collaboration), JCAP \textbf{04} 038 (2017)

\bibitem{Auger_CF_2023}
A. Abdul Halim et al. (The Pierre Auger Collaboration), JCAP \textbf{05} 024 (2023)

\bibitem{Auger_CFAD_2023}
A. Abdul Halim et al. (The Pierre Auger Collaboration), submitted to JCAP, arXiv:2305.16693

\bibitem{CRPropa}
R. Alves Batista et al., JCAP \textbf{1605} 038 (2016)

\bibitem{Gilmore}
R.~Gilmore et al., MNRS \textbf{422} 3189 (2012)

\bibitem{TALYS} 
A.~Koning and D.~Rochman, Nucl. Data Sheets \textbf{113} 2841 (2012)

\bibitem{healpy} A. Zonca et al., J. Open Source Softw. \textbf{4} 35 (2019)

\bibitem{JF12a} R. Jansson and G. R. Farrar, ApJ \textbf{757} 14 (2012)

\bibitem{JF12b} R. Jansson and G. R. Farrar, ApJL \textbf{761} L11 (2012)

\bibitem{Ding_ICRC_2019} N. Globus, C. Ding and G. R. Farrar, PoS \textbf{ICRC2019} 243 (2019)





\bibitem{Almeida_dipole_2021}
R. de Almeida for the Pierre Auger Collaboration, Pos \textbf{ICRC2021} 335 (2021)

\bibitem{Auger_E_2020}
A. Aab et al. (The Pierre Auger Collaboration), PRD \textbf{102}, 062005 (2020)

\bibitem{EPOS}
T. Pierog, I. Karpenko, J.M. Katzy, E. Yatsenko and K. Werner, PRD \textbf{92} 034906 (2013) 

\bibitem{Yushkov_xmax_2019}
A. Yushkov for the Pierre Auger Collaboration, Pos \textbf{ICRC2019} 482 (2019)

\bibitem{UF23}
M. Unger and G. R. Farrar, in preparation

\bibitem{Mayotte_ICRC_2021}
E. Mayotte for the Pierre Auger Collaboration, Pos \textbf{ICRC2021} 321 (2021)

\bibitem{Mayotte_ICRC_2023}
E. Mayotte for the Pierre Auger Collaboration, Pos \textbf{ICRC2023} 365 (2023)

\bibitem{CF3}
D. Pomarède, R. B. Tully, R. Graziani, H. Courtois, Y. Hoffman, and J. Lezmy, ApJ \textbf{897} 133 (2020)




\end{thebibliography}
\end{document}